\documentclass{nature}
\usepackage{graphicx}
\usepackage{amssymb}
\usepackage{float}
\usepackage{bm}
\usepackage{array}
\usepackage{amsmath}
\usepackage{booktabs} 
\usepackage{multicol}

\graphicspath{{Pic/}}

\title{Significant Impact of Rossby Waves on Air Pollution Detected by Network Analysis}

\author{Yongwen Zhang$^{1,2,3}$, Jingfang Fan$^{2,4}$, Xiaosong Chen$^{5}$, Yosef Ashkenazy$^1$ \& Shlomo Havlin$^2$}

\begin{document}

\maketitle

\begin{affiliations}
 \item Department of Solar Energy and Environmental Physics, The Jacob Blaustein Institutes for Desert Research, Ben-Gurion University of the Negev, Midreshet Ben-Gurion 84990, Israel;
 \item Department of Physics, Bar-Ilan University, Ramat Gan 52900, Israel;
 \item Data Science Research Center, Faculty of Science, Kunming University of Science and Technology, Kunming 650500, Yunnan, China;
 \item Potsdam Institute for Climate Impact Research, 14412 Potsdam, Germany;
 \item  School of Systems Science, Beijing Normal University, Beijing 100875, China. 
 \end{affiliations}

\begin{abstract}
 Air pollution is associated with human diseases and has been found to be related to premature mortality\cite{Heft-Neal, Lelieveld}. In response, environmental policies have been adopted in many countries, to decrease anthropogenic air pollution for the improvement of long-term air quality, since most air pollutant sources are anthropogenic. However, air pollution fluctuations have been found to strongly depend on the weather dynamics\cite{Tai,cai2017weather}. This raises a fundamental question: What are the significant atmospheric processes that affect the local daily variability of air pollution? For this purpose, we develop here a multi-layered network analysis to detect the interlinks between the geopotential height of the upper air ($\sim$ 5 km) and surface air pollution in both China and the USA. We find that Rossby waves significantly affect air pollution fluctuations through the development of cyclone and anticyclone systems, and further affect the local stability of the air and the winds. The significant impacts of Rossby waves on air pollution are found to underlie most of the daily fluctuations in air pollution. Thus, the impact of Rossby waves on human life is greater than previously assumed\cite{Boers2019}. The rapid warming of the Arctic could slow down Rossby waves\cite{Francis2012}, thus increasing human health risks. Our method can help to determine the risk assessment of such extreme events and can improve potential predictability.
\end{abstract}

During the last two decades, complex network approaches have been applied to a wide range of disciplines for studying complex systems\cite{Barabasi1999a,Cohen2010,Newman2018}. Network approaches have been found useful for detecting and better understanding the statistical and dynamical features of complex systems\cite{Helbing}. In climate networks, geographical locations or observation sites can be regarded as nodes, and their interactions can be quantified as links. The strength of the links is determined by the level of similarity between climate variables (such as temperature, pressure and precipitation) determined, for example, by cross-correlations. Climate network techniques have been successfully applied to improve our understanding of climate phenomena such as the El Ni$\tilde{n}$o/Southern Oscillation (ENSO), the North Atlantic Oscillation (NAO) and Rossby waves\cite{Tsonis2008a,Donges2009d,Guez2012,Wang2013b}. 
Climate networks have also been used to forecast extreme climate events\cite{Ludescher2014a,Boers2014}. Furthermore, network approaches can provide a new way to detect and evaluate the impacts of past and future climate change\cite{Fan2}. A global rainfall teleconnection pattern related to upper-level Rossby waves has been recently revealed using complex network analysis\cite{Boers2019}.  

Outdoor air pollution plays an important role in widespread environmental and health problems. Life expectancy is closely related to air quality\cite{Heft-Neal}. Fine particulate matter, PM$_{2.5}$, of an aerodynamic diameter of $2.5$ $\mu m$ or less, is a very harmful air pollutant to human health and has been associated with lung cancer and heart diseases\cite{Burnett2014}. It is responsible for most of the global mortality caused by outdoor air pollution\cite{Cohen2017}. This is especially significant in developing countries due to their rapid economic development and the corresponding rapid increase in anthropogenic greenhouse gas emissions. In recent years, China has experienced increasingly frequent and persistent haze events caused by high PM$_{2.5}$ concentrations\cite{Zhang2015}. Millions of premature deaths have been associated with outdoor air pollution in China. Even in a developed country like the USA, there are still thousands of premature deaths every year due to outdoor air pollution\cite{Lelieveld}. Air pollution can also potentially affect climate systems\cite{Yohanan}. The major sources of particulate matter pollution are both anthropogenic and natural, including traffic emissions, industrial production, biomass burning, agricultural activities and dust\cite{Zhang2015}.

Atmospheric circulations can cause particulate matter to be transported thousands of kilometers\cite{Liu}. Goods and services produced in one region for consumption in another (inter-regional trade) can also substantially contribute to the local particulate matter concentration\cite{Zhang}. Quantifying the contributions from the various sources has become a central issue in air pollution studies\cite{Huyong}. However, the variability of particulate matter pollution does not only depend on the sources. Previous studies suggested that air pollution variability strongly depends on meteorological conditions\cite{Tai,cai2017weather}. Extreme air pollution events are usually associated with weak surface wind, a shallow boundary layer, temperature inversion and more. Such weather conditions could be linked to a common atmospheric mechanism in the upper air. For example, it was suggested that most hazy days in the North China Plain are associated with a local anticyclone in the mid-troposphere\cite{Chen2018}. Thus, it is important to identify and quantify the connections between air pollution and the atmospheric dynamics of the mid and upper air levels.

Here, we perform a multi-layer and multi-variable network analysis and identify the interlinks between the 500 hPa geopotential height and surface air pollution; this is aimed at quantifying the influence of the upper air dynamics on the temporal variability of surface air pollution. We mainly consider the PM$_{2.5}$ concentrations in three regions: China, and the western and eastern USA. The detrended daily observation grids of PM$_{2.5}$ and the geopotential heights are used to construct networks (Methods). We identify two multi-layered networks, one with dominant anti-correlated interlinks (between geopotential heights and pollution) and one with positively correlated interlinks, corresponding to negative and positive weights, respectively (Methods). Upper air critical regions are quantified by measuring the out-degree of each upper air node associated with the total weights of the significant outgoing interlinks to the ground nodes of pollution. Furthermore, affected regions on the ground are quantified by the in-degree of each ground node associated with the total weights of the significant incoming interlinks from the upper air nodes.


\begin{figure}
\centering
\includegraphics[scale=0.5]{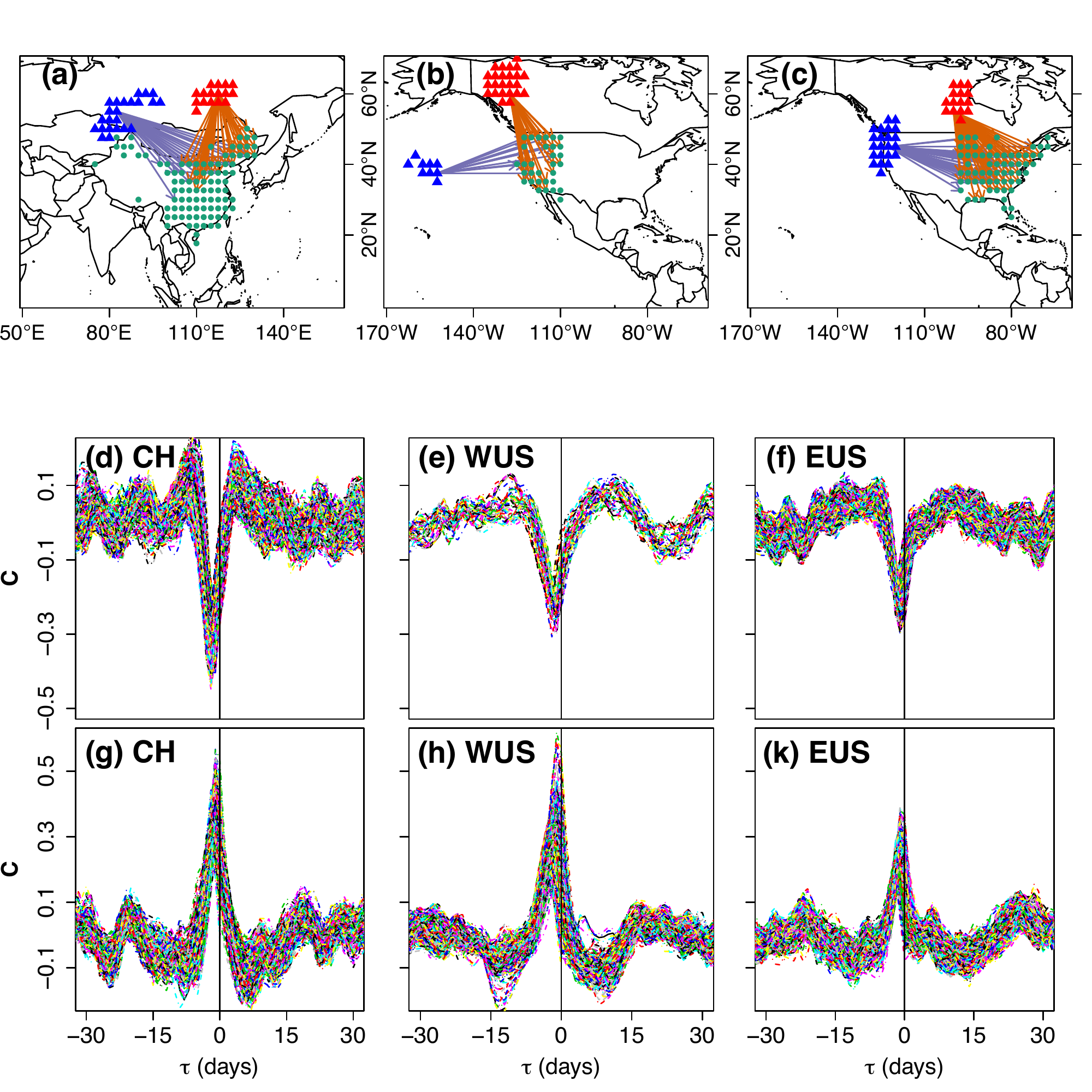}
\caption{(a-c) The multi-layered maps. Blue triangles represent the strongest 10\% negative (negatively correlated) out-degree nodes (outgoing from the upper level nodes), and red triangles are the strongest 10\% positive out-degree nodes in the upper air. The interlinks (blue and red arrows) are between the 500 hPa geopotential height and the surface PM$_{2.5}$ concentration (green points). To avoid overlapping, all red triangles are moved $15^{\circ}$ to the north. The lower six panels depict the cross-correlation functions of the negative and positive networks' interlinks (different interlinks with different colors) for (d), (g) China, (e), (h) the western, and (f), (k) the eastern USA, respectively.}     
\label{example}
\end{figure}

Figure \ref{example}a-c presents the strongest 10\% out-degree nodes of the positively correlated and negatively correlated multi-layered networks composed of the 500 hPa geopotential height and the surface air PM$_{2.5}$ concentration. The interlinks’ direction is toward the air pollution nodes at the surface. The weights of the interlinks are obtained from the cross-correlation functions (shown in Fig. \ref{example}d-k). The blue triangles at 500 hPa reveal strong anti-correlations with the air pollution nodes. Significant negative peaks are observed for all three regions (Fig. \ref{example}d-f). Most of the negative peaks are found for time lags of $-1$ or $-2$ days, indicating that changes of the 500 hPa geopotential height occur one or two days before the correspondingly opposite changes in the surface PM$_{2.5}$ concentration. In contrast, the interlinks of the positive multi-layered network between the 500 hPa geopotential height and the surface air PM$_{2.5}$ concentration indicate that the correlation functions have strong positive peaks with a $-1$ day time delay (Fig. \ref{example}g-k), meaning that a high geopotential height is observed $\sim$1 day before a high level of surface PM$_{2.5}$ concentration. In the western and eastern USA, the above interlinks are connected to most of the ground nodes. In China, the interlinks are mainly connected to northeastern China, a region that has been severely and frequently exposed to PM$_{2.5}$ pollution in recent years\cite{Zhang2015}. Our results indicate that these locations are most influenced by the upper air. We obtain similar results for the PM$_{10}$ concentration (Supplementary Fig. 1).
     
\begin{figure}
\centering
\includegraphics[scale=0.5]{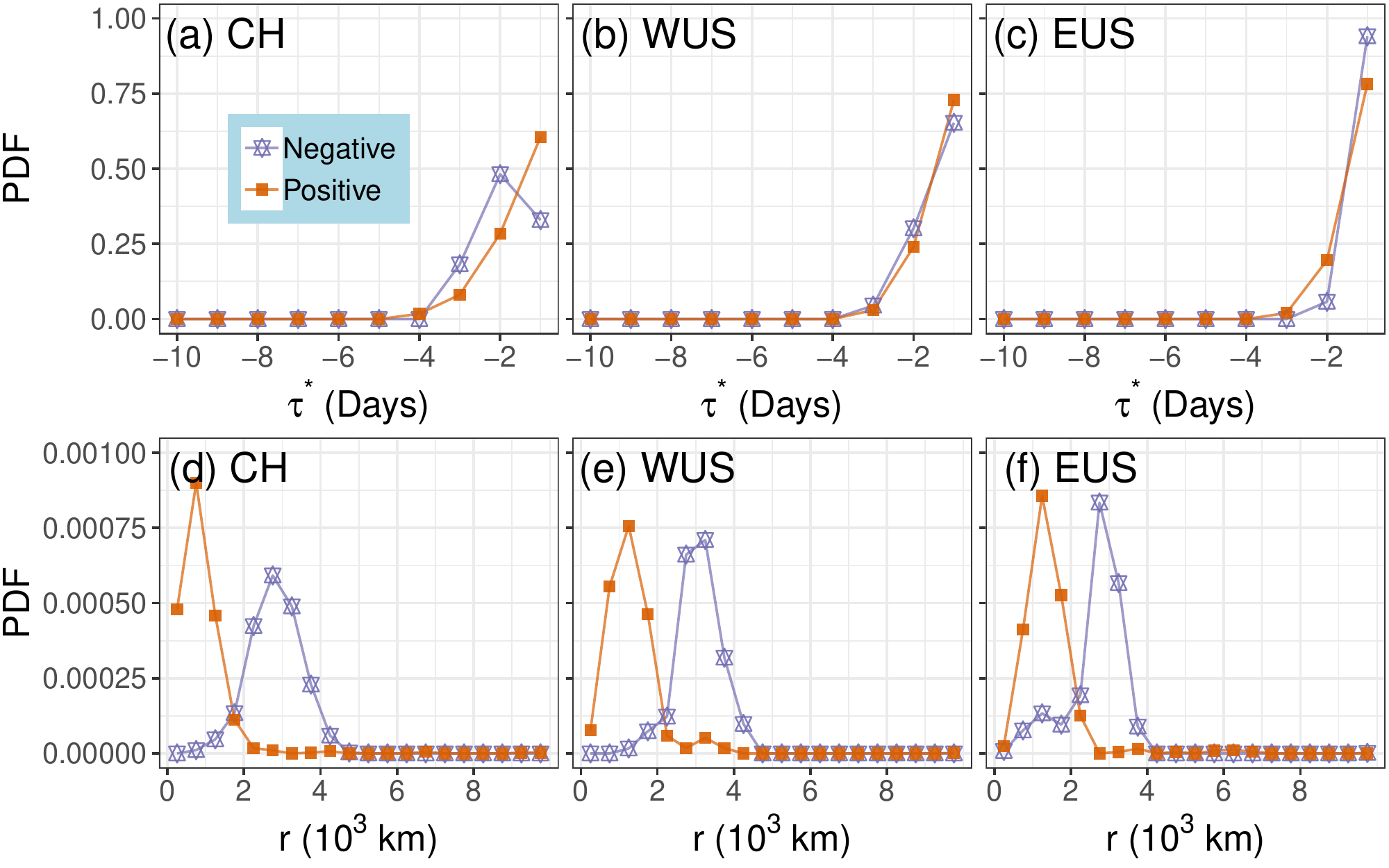}
\caption{Interlinks statistics: (a), (b), (c) PDF of the time lag $\tau^*<0$ of the interlinks, and (d), (e), (f) PDF of the geographical length of the interlinks for China, and the western and eastern USA, respectively.} 
\label{Wtd}
\end{figure}

Figure \ref{Wtd}a-c shows the probability distribution function (PDF) of the time lag of the interlinks for China, and the western and eastern USA, respectively. For China (Fig. \ref{Wtd}a), the maximum probability of the time lag is at $-2$ days for the negative correlations. All other distributions show the maximum values at $-1$ days. The PDFs in Fig. \ref{Wtd}a-c, show a decay to zero after approximately $4$ days, indicating that the maximal influence time of the geopotential height on the PM$_{2.5}$ level is four days in China and about three days in the USA. This difference could be associated to the faster jet stream in the USA. We also calculate the PDF of the geographical lengths of the interlinks (shown in Fig. \ref{Wtd}d-f). Two peaks are found around 1,000 km and 3,000 km for the positive and negative interlinks, respectively.
Based on refs.\cite{Chang1999, Wang2013b}, we suggest that the distributions of the distances are related to atmospheric Rossby waves whose wavelengths vary from around 5,000 km to 7,000 km. The longer distances of the negative interlinks ($\sim$3000km) seem to correspond to half a wavelength of the Rossby wave.
The peak in the short distances of the positive interlinks ($\sim$1,000 km) could be associated with a typical cyclonic scale within the Rossby wave pattern. 

\begin{figure}
\centering
\includegraphics[scale=0.35]{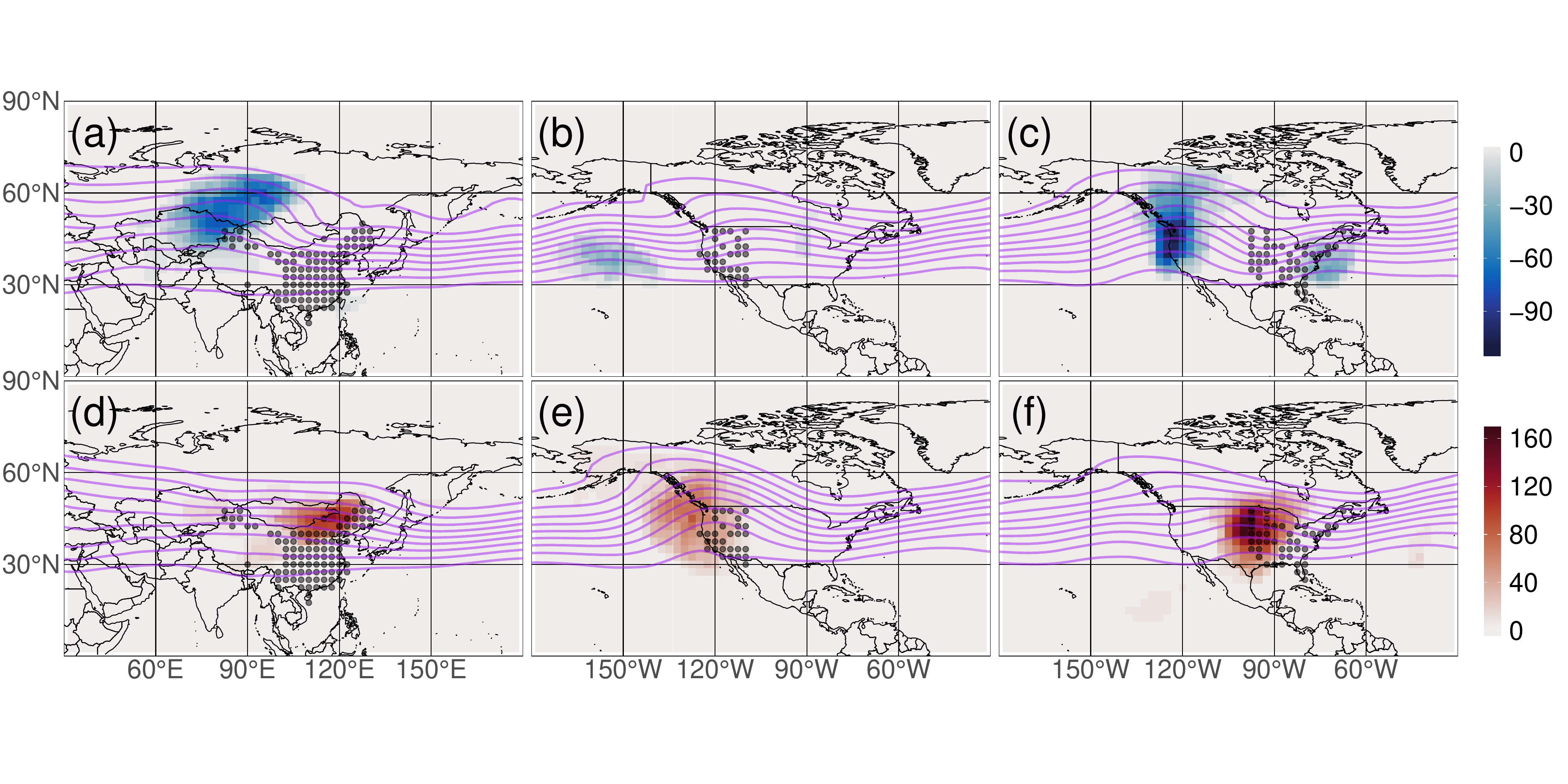}
\caption{Maps of the 500 hPa geopotential height: Distributions of (a-c) negative and (d-f) positive out-degrees in the upper air for China, and the western and eastern USA. Pink contours depict the mean geopotential height of two extreme cases of days with (a-c) extremely low and (d-f) extremely high PM$_{2.5}$ concentrations. The values of the contours increase from north to south within a range of 5300 to 5700 m. Grey circles show the locations of the surface PM$_{2.5}$ nodes.} 
\label{out}
\end{figure}

To further support the association of the observed network pattern with Rossby waves, we compare the location of out-degree nodes with the contours of the geopotential height at 500 hPa whose meandering shapes can generally describe the amplitude and phase of a Rossby wave (i.e., corresponding to a ridge or trough\cite{Wirth2018}). We consider the mean geopotential height of two extreme cases of days that are one day before extremely low (the lowest 10\% of days in each month) and extremely high (the top 10\% of days in each month) PM$_{2.5}$ concentration days over northeastern China, and the western and eastern USA. Figure \ref{out}a-c shows the contours of the mean geopotential height for the extremely low air pollution cases, and Figure \ref{out}d-f shows the extremely high air pollution cases in the above three regimes. The localized nodes of the strongest negative out-degree links (over China) are found to be in Russia (Fig. \ref{out}a). Interestingly, in the same region, the contours of geopotential height form a ridge for the low air pollution days. A trough is also found for the low air pollution days in the east, associated with the red positive nodes of out-degrees, shown in Figure \ref{out}d. During the high air pollution days, the contours show a small ridge in the vicinity of the red positive cluster (Fig. \ref{out}d), and the ridge in Russia disappears. The USA networks exhibit similar results. Yet, the air pollution over the western and eastern USA is affected by different (upper air) ridges and troughs (Figs. \ref{out}b, c, e and f), leading to the different daily variations of air pollution in the western and the eastern USA\cite{Tai}. The out-degrees of the networks are generally in accordance with the ridges and troughs (associated with Rossby waves), suggesting that these are important regions in the upper air that influence the surface air pollution. Note that we obtained similar patterns when using the geopotential heights of 300 hPa and 700 hPa (Supplementary Figs. 4 and 6), indicating that the above mentioned patterns are robust and related to large-scale atmospheric circulation. The in-degree patterns are also shown (in Supplementary Figs. 3, 5 and 7).

\begin{figure}
\centering
\includegraphics[scale=0.5]{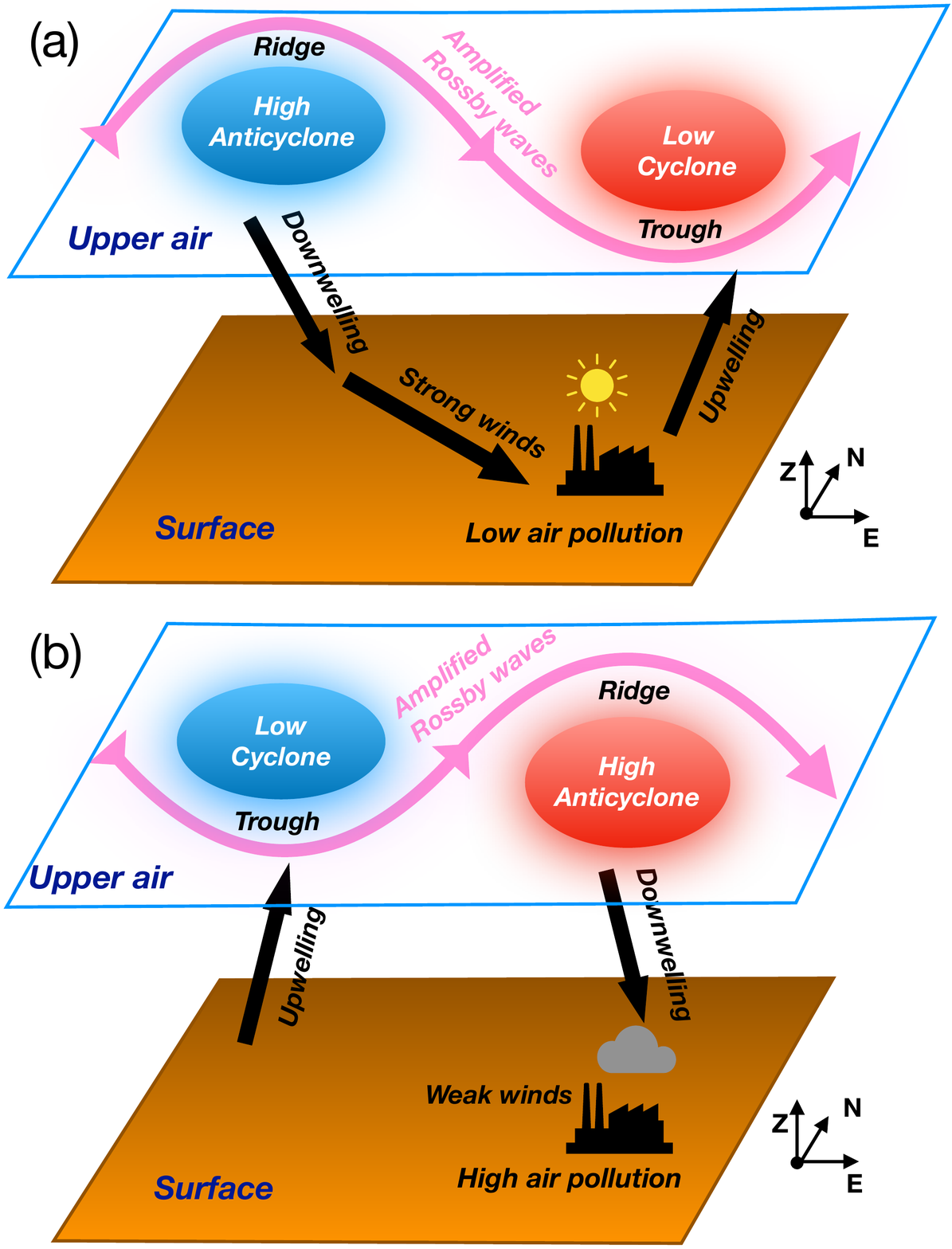}
\caption{Schematic representation of the atmospheric mechanism caused by Rossby waves. (a) and (b) correspond to low and high air pollution conditions respectively. See the text for further details. }
\label{msk}
\end{figure}

Next, we propose a mechanism for the impact of Rossby waves on surface air pollution. Two extreme phases are considered here and depicted schematically in Figure \ref{msk}. The blue and red colors represent the negative and positive out-degree clusters, respectively. The negative ones are usually located to the west of the air pollution nodes. This can be understood by the high geopotential height that precedes the low pollution events and since the winds blow from west to east in jet streams. The flow often meanders to the north and south to form ridges and troughs caused by Rossby waves. For the phase of the Rossby wave depicted in Figure \ref{msk}a, an anticyclone emerges on the (left) ridge, leading to downwelling, while, at the same time, a cyclone develops on the (right) trough near the pollution areas, leading to upwelling. In addition, a high pressure system develops at the surface, to the west of the air pollution nodes (Supplementary Fig. 8). The downwelling air to the west of the pollution nodes and the upwelling air in the vicinity of the pollution nodes intensify the westerly winds, which effectively disperse the air pollutants.
Moreover, the cyclone (and its upwelling air) near the pollution nodes carries pollutants to the upper air, decreasing surface pollutants. For the opposite phase of the Rossby wave depicted in Figure \ref{msk}b, the anticyclone and the cyclone switch their locations. Downwelling changes to upwelling in the west, which diverts part of the surface westerly wind upward, hence weakening the surface winds (Supplementary Fig. 8). In the vicinity of the pollution location, the (eastern) anticyclone causes the upper air to downwell. As a result, a temperature inversion layer forms, and the boundary layer is low, leading to a stable air column that traps the pollutants close to the surface, hence leading to severe pollution. Thus, the upper air out-degree field of the multi-layered network provides important information regarding the daily variability of surface air pollution. Potentially, the time delay of the links and their strength may also be useful as a predictive tool. 

\begin{figure}
\centering
\includegraphics[scale=0.5]{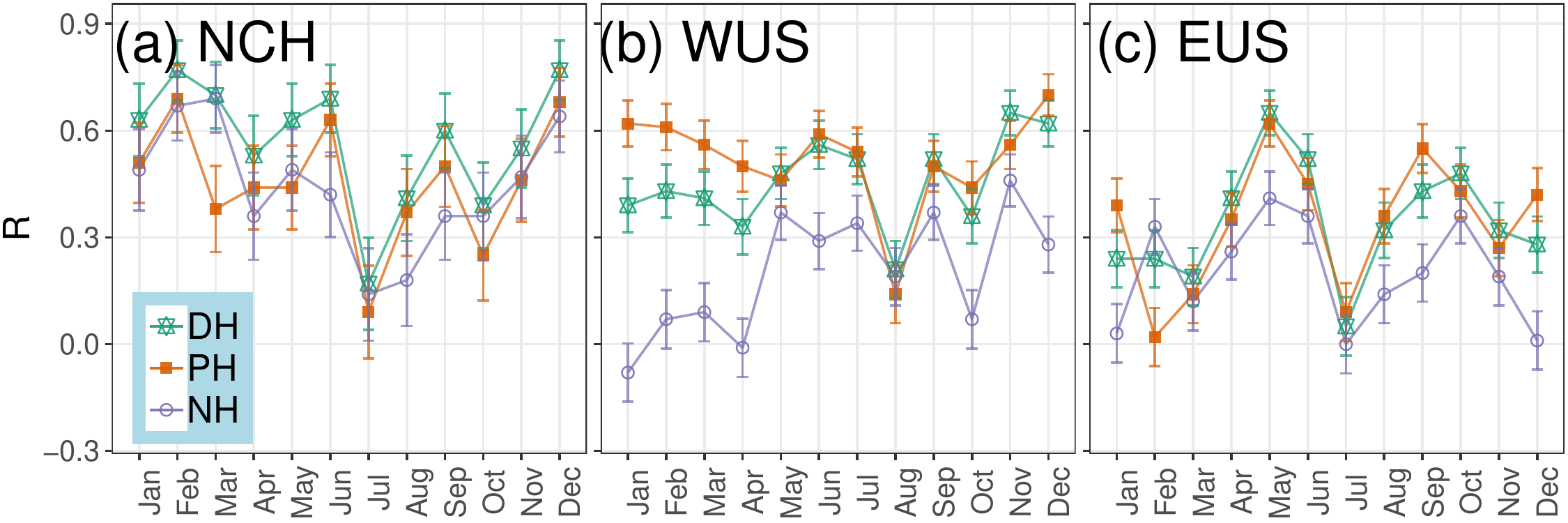}
\caption{(Color online) Pearson correlation coefficient $R$ between the geopotential heights $DH$, $PH$, $NH$ and the daily averaged PM$_{2.5}$ concentration with a one-day delay over (a) northeastern China, (b) the western USA and (c) the eastern USA , for different months in the year. The R value of NH is multiplied by $-1$. The standard error bar is given by $\sqrt{\frac{1-R^2}{n-2}}$.}
\label{PR}
\end{figure}

We next test the monthly effects on the interactions between the geopotential heights and the air pollution. We denote the daily averaged negative and positive geopotential heights (discussed above) as ${\rm NH}$ and ${\rm PH}$, respectively, in the upper air nodes (the blue and red triangles shown in Figure \ref{example}). We also denote the difference between the geopotential height as ${\rm DH}={\rm PH}-{\rm NH}$ in order to test the potential predictive power when considering both the negative and positive patterns. The specific scatter plots and the corresponding correlation curves of the daily averaged PM$_{2.5}$ concentration (with a one-day delay) versus DH are shown in Supplementary Figures 9-11. Most of the months display good linear relations. Figure \ref{PR} shows their Pearson correlation coefficient $R$. The $R$ values of DH are almost always the highest, especially over northeastern China where the $R$ value is usually larger than 0.5. For several months, the $R$ values of DH over the western USA are similar to PH; this can be attributed to the weak anti-correlation and predominant positive correlation patterns. We note that the $R$ values during the summer (July and August) are significantly lower than in other months for all regions. This is probably since the jet stream and the winds are much weaker during the summer months, resulting in weaker interactions between the upper air and ground pollution. In general, when considering both the negative and positive patterns, taking the difference between them (DH) can provide a better predictive tool.

Indeed, most sources of air pollutants are anthropogenic. Yet, the variability of air pollution strongly depends on atmospheric processes.
Here, we have found that the response of the PM$_{2.5}$ pollution to the geopotential height of 500 hPa is significant. We identify the corresponding upper air regions using network analysis and quantify their influence; there are two localized areas (western and eastern) in both China and the USA. The western region is anti-correlated with the surface air pollution nodes, while, in contrast, the eastern region (which is closer to the pollution region) is positively correlated with the upper air pollution nodes. We suggest that the underlying mechanism is related to Rossby waves that are associated with meanders of the mid-latitude jet stream and with ridges and troughs of the upper air geopotential height, leading to high (anticyclonic) and low (cyclonic) pressure systems. Thus, the amplitudes and phases of Rossby waves substantially affect air pollution on the ground. Furthermore, the difference of geopotential height between the two regions has a potential predictive power.

Earlier studies\cite{Petoukhov2013,Screen2014} demonstrated the statistical connection between high-amplitude Rossby waves and extreme weather events on the ground. These include heat and cold waves that occurred in the Northern Hemisphere mid-latitudes during the past decade. Here, we show that pronounced Rossby waves not only cause extreme weather events but also lead to stronger fluctuations of air pollution. The two extreme phases of Rossby waves lead to low and high surface pollution. Since the Arctic sea ice extent is rapidly declining, the warming in this region is occurring faster than in the rest of the Northern Hemisphere, leading to a decreasing meridional temperature gradient that could cause the jet stream and winds to weaken\cite{Archer2008}. Therefore, a slower eastward propagation of Rossby waves may develop\cite{Francis2012}, which could be associated with more prolonged extreme weather events and may lead in the future, as suggested by the present study, to more prolonged air pollution events\cite{cai2017weather}.

The interlinks of the multi-layered network can be used to quantify the relationship between the upper troposphere and the surface pollution. We find here indications that the amplitudes and phases of Rossby waves influence the surface air pollution. Our study not only provides significant spatial patterns at different levels but also shows time shifts between the levels that can help to determine the risk assessment of extreme events in space and improve potential predictability. Our method can also be used to study other interactions of weather patterns between the atmosphere and the surface.

\begin{methods}

\subsection{Data pre-processing}
The daily average surface mass concentrations of PM$_{2.5}$ in $364$ cities across China from 2015 to 2017 are measured by the Chinese Ministry of Environmental Protection (http://113.108.142.147:20035/emcpublish/). Most of the cities are in eastern China. The daily mean surface PM$_{2.5}$ concentrations measured in $\sim$1000 sites in the USA from 2013 to 2017 were provided by the Environmental Protection Agency (EPA) Air Quality System (AQS) (https://\\www.epa.gov/outdoor-air-quality-data/). Since the zonal extent of the USA is large enough, we considered two regions: the western and eastern USA. The air pollution sites are evenly distributed, and we interpolated the data onto $2.5^{\circ} \times 2.5^{\circ}$ grids. The grid points that included more than 5\% missing values were omitted, resulting in $N_A = 95$, $31$ and $63$ grid points for China, and the eastern and western USA, respectively. The geopotential height data are from the global ERA-interim reanalysis ($2.5^{\circ} \times 2.5^{\circ}$) of the European Center for Medium-Range Weather Forecasts (ECMWF)\cite{Dee2011}. To filter out trivial seasonal effects, we first detrended the original data by subtracting the corresponding centered 30-day ``moving'' means and then normalized by the standard deviation of these 30 days.

\subsection{Network construction}
The detrended daily surface PM$_{2.5}$ concentration of grid point $i$ and the geopotential height (500 hPa) of grid point $j$ are represented by time series $A_{i} (t)$ and $M_{j}(t)$ during a time period $T$. The fluctuation series are  $\delta A_{i} (t)= A_{i} (t)-\left\langle A_{i} \right\rangle$ and $\delta M_{j} (t)= M_{j} (t)-\left\langle M_{j} \right\rangle$, where $\left\langle A_{i} \right\rangle$ and $\left\langle M_{i} \right\rangle$ are temporal averages for the period $T$. 
The cross-correlation function between the two variables for grids $i$ and $j$ is defined as
\begin{equation}
C_{A_i, M_j}(\tau)=\frac{\left\langle \delta A_{i}(t-\tau) \cdot \delta M_{j}(t) \right\rangle}{\sqrt{\left\langle \left[ \delta A_{i} (t-\tau) \right]^{2} \right\rangle} \cdot \sqrt{\left\langle \left[ \delta M_{j} (t)\right]^{2} \right\rangle}}\;,
\label{corr}
\end{equation}
where the time lag $\tau$ is constrained to be within $-\tau_{max} \le \tau \le \tau_{max}$ where $\tau_{max}=90$ days.
We identify the maximal absolute value of $C_{A_i, M_j}(\tau)$ and denote the corresponding time lag as $\tau^*$ (Fig. \ref{example}d-k). The direction of correlation (link) $C_{A_i, M_j}(\tau^*)$ is from $A_i$ to $M_j$ when $\tau^* > 0$ and from $M_j$ to $A_i$ when $\tau^*< 0$. The direction is undefined for $\tau^*=0$.  

Based on ref.\cite{Yamasaki2008b,Fan} we quantify the strength of the correlations using 
\begin{equation}
W_{A_i, M_j}=\frac{C_{A_i, M_j}(\tau^*)-mean(C_{A_i, M_j}(\tau))}{std(C_{A_i, M_j}(\tau))}\;,
\label{W}
\end{equation}
where ``mean'' and ``std'' represent the mean and standard deviation of the cross-correlation function, respectively.

We choose a threshold of $W$ to be $\Delta=4.5$, for which the link strength is considered to be significant; this is based on shuffled data significance tests. The weight $W$ as a function of $\tau^*$ and distance $r$ also supports the reliability of the threshold (Supplementary Fig. 2). Because the negative and positive weights have different properties, we introduce two non-square matrices
$\bm{X}^p$ and $\bm{X}^n$ to describe the interlinks of the positively and negatively correlated multi-layer networks, respectively.
The element $X^p_{i, j}$ will equal one, if $ W_{A_i, M_j} \geqslant \Delta$ and $\tau^*< 0$ otherwise $X^p_{i, j}$ is zero. In the same way, the element $X^n_{i, j}$ will equal one, if $W_{A_i, M_j} \leqslant -\Delta$ and $\tau^*< 0$; otherwise $X^n_{i, j}$ is zero.
 
 In our networks, the grid sites are regarded as nodes. Then, the negative and positive out-degrees of node (grid) $j$ for the geopotential height are calculated by
 \begin{equation}
 \begin{split} 
O_{M_j}^n=\sum_{i=1}^{N_A} X^n_{i, j}W_{A_i, M_j}\;,
\\
O_{M_j}^p=\sum_{i=1}^{N_A} X^p_{i, j}W_{A_i, M_j}\;.
\end{split}
\label{oDeg}
\end{equation}

For the air pollution, we have negative and positive in-degrees of node (grid) $i$: 
\begin{equation}
 \begin{split} 
I_{A_i}^n=\sum_{j=1}^{N_M} X^n_{i, j}W_{A_i, M_j}\;,
\\
I_{A_i}^p=\sum_{j=1}^{N_M} X^p_{ij}W_{A_i, M_j}\;.
\end{split}
\label{iDeg}
\end{equation}

Obviously, the outgoing links of the geopotential height are the same as the incoming links of the air pollution in a multi-layer network. In order to describe the significance of the nodes in the two layers, the out- and in-degrees measure the number of strong outgoing and incoming links from and to the geopotential height nodes and the air pollution nodes, respectively.



\end{methods}




\begin{addendum}
 \item We thank Dean Chen and Liu Wenqi for helpful discussions. We thank the Italian Ministry of foreign affairs and international cooperation, and the Israeli Ministry of science, technology, and space; the Israel Science Foundation, ONR, Japan Science Foundation, BSF-NSF, ARO, the EU H2020 project RISE, and DTRA (Grant no. HDTRA-1-10-1-0014) for financial support. We also acknowledge the data resources provided by the Ministry of Environmental Protection of China and United States Environmental Protection Agency.
 \item[Competing Interests] The authors declare that they have no
competing financial interests.
 \item[Correspondence] Correspondence and requests for materials
should be addressed to Yongwen Zhang (email: zhangyongwen77@gmail.com).
\end{addendum}


\end{document}